# Planets and Dark Energy

Carl H. Gibson [1,2]

[1] University of California San Diego, La Jolla, CA 92093-0411, USA
[2] cgibson@ucsd.edu, http://sdcc3.ucsd.edu/~ir118

and

Rudolph E. Schild[3,4]

[3] Center for Astrophysics, 60 Garden Street, Cambridge, MA 02138, USA
[4] rschild@cfa.harvard.edu

**Abstract:** Self-gravitational fluid mechanical methods termed hydro-gravitational-dynamics (HGD) predict plasma fragmentation 0.03 Myr after the turbulent big bang to form protosuperclustervoids, turbulent protosuperclusters, and protogalaxies at the 0.3 Myr transition from plasma to gas. Linear protogalaxyclusters fragment at 0.003 Mpc viscous-inertial scales along turbulent vortex lines or in spirals, as observed. The plasma protogalaxies fragment on transition into white-hot planet-mass gas clouds (PFPs) in million-solar-mass clumps (PGCs) that become globular-star-clusters (GCs) from tidal forces or dark matter (PGCs) by freezing and diffusion into 0.3 Mpc halos with 97% of the galaxy mass. The weakly collisional non-baryonic dark matter diffuses to > Mpc scales and fragments to form galaxy cluster halos. Stars and larger planets form by binary mergers of the trillion PFPs per PGC, mostly on 0.03 Mpc galaxy accretion disks. Stars deaths depend on rates of planet accretion and internal star mixing. Moderate accretion rates produce white dwarfs that evaporate surrounding gas planets by spin-radiation to form planetary nebulae before Supernova Ia events, dimming some events to give systematic distance errors, the dark energy hypothesis, and overestimates of the universe age.

## 1. Introduction

Dimness of supernovae Ia (SNe Ia) events for redshift values $0.01 < z < 2$ have been interpreted as an accelerating rather than decelerating expansion rate for the universe [1-3]. The acceleration is attributed to mysterious antigravity effects of "dark energy" and a cosmological constant $\Lambda$. Dimming is observed at all frequencies by about 30%, with large scatter attributed to uncertainty in the SNe Ia models. Hubble Space Telescope Advanced Camera for Surveys (HST/ACS) images have such high signal to noise ratios that both the scatter and the dimming are statistically significant over the full range of z values. Bright SNe Ia observed for $z \geq 0.46$ exclude "uniform grey dust" systematic errors, supporting flat-universe deceleration until the recent "cosmic jerk" to acceleration for $z \leq 0.46$. The "dark energy" interpretation is a consequence of the commonly accepted $\Lambda$-cold-dark-matter ($\Lambda$CDM) cosmological theory. However, $\Lambda$CDM theory is fluid mechanically untenable. It is collisionless and frictionless, neglecting effects of viscosity, turbulence[1], diffusion, fossil turbulence and fossil turbulence waves. Hydro-gravitational-dynamics (HGD) is the application of modern fluid mechanics to cosmology [4-13]. Primordial planets at $z = 1100$, $t = 10^{13}$ s predicted by HGD are the source of all stars and the dark matter of galaxies [4,14]. Because dying stars may be dimmed by gas planets evaporated near the star, planets provide an alternative to the revolutionary rejection of standard physics required by dark energy. New physical laws are not required by HGD but $\Lambda$CDM must be discarded. The choice is between planets and dark energy.

In the following Section 2 we discuss the theories of gravitational structure formation, followed by a review of the observational data in Section 3, a Discussion of Results in Section 4 and finally some Conclusions in Section 5.

---

[1] Turbulence is defined as an eddy-like state of fluid motion where the inertial-vortex forces of the eddies are larger than any other forces that tend to damp the eddies out. Turbulence by this definition always cascades from small scales to large, starting at the viscous-inertial-vortex Kolmogorov scale at a universal critical Reynolds number. Irrotational flows cannot be turbulent by definition. In self-gravitational fluids such as stars and planets, turbulence is fossilized in the radial direction by buoyancy forces. Radial transport is dominated by fossil turbulence waves and secondary (zombie) turbulence and zombie turbulence waves in a beamed, radial, hydrodynamic-maser action.

## 2. Theory

The Jeans 1902 linear theory of acoustic gravitational instability [15] predicts only one criterion for structure formation. Fluctuations of density are unstable at length scales larger than the Jeans length $L_J = V_s / (\rho G)^{1/2}$ but stable for smaller scales, where $V_s$ is the speed of sound, $\rho$ is density, $G$ is Newton's gravitational constant, and $(\rho G)^{-1/2}$ is the gravitational free fall time. Because the speed of sound in the plasma epoch after the big bang is nearly the speed of light, the Jeans scale for the plasma is always larger than the scale of causal connection $ct$, where $c$ is the speed of light and $t$ is the time since the big bang, so no gravitational structures can form in the plasma. Jeans' 1902 fluid mechanical model [15] is the basis of ΛCDM, where an unknown form of collisionless non-baryonic dark matter (NBDM) is assumed to condense because it is somehow created cold so its sound speed and Jeans length are assumed smaller than $ct$. The NBDM clumps cluster hierarchically and somehow stick together to form potential wells in ΛCDMHC models [1, 2] with dark energy. CDM clumps have been sought but not observed, as predicted by HGD.

It is not true that the primordial plasma is collisionless and it is not true that gravitational instability is linear. Gravitational instability is intrinsically nonlinear and absolute. All fluid density fluctuations are unstable to gravity, and in the plasma epoch $10^{11} \le t \le 10^{13}$ s structures will form by gravitational forces at all scales less than $ct$ unless prevented by diffusion, viscous forces or turbulence forces, as shown in Figure 1 [11]. The kinematic viscosity of the plasma is determined by photon collisions with free electrons that drag their protons and alpha particles with them [5]. When the Schwarz viscous-gravitational scale $L_{SV}$ becomes less than the horizon scale $L_H = ct$ then gravitational structure formation begins.

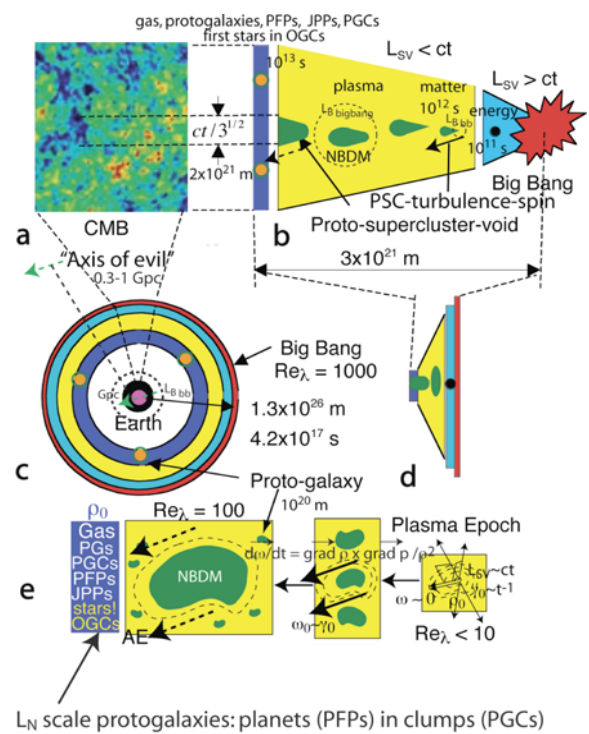

Fig. 1. Formation of gravitational structures according to hydro-gravitational-dynamics (HGD). The entire baryonic plasma universe fragments at the Schwarz viscous scale to form planets in Jeans mass clumps in Nomura geometry [9] protogalaxies at the plasma to gas transition. Batchelor scale fossil big bang turbulence density gradients produce "axis of evil" quasar, galaxy and galaxy cluster spin alignments to Gpc scales [11, 22].

The density and rate-of-strain of the plasma at transition to gas at 300,000 years ($10^{13}$ s) are preserved as fossils of the time of first structure at 30,000 years ($10^{12}$ s), as shown in Fig. 1e. The plasma turbulence is weak at transition, so the Schwarz viscous scale $L_{SV} = (\gamma \nu / \rho G)^{1/2}$ and Schwarz turbulence scale $L_{ST} = \varepsilon^{1/2} / (\rho G)^{3/4}$ are nearly equal, where $\gamma$ is the rate-of-strain, $\nu$ is the kinematic viscosity, $\varepsilon$ is the viscous dissipation rate and $\rho$ is the density. Because the temperature, density, rate-of-strain, composition and thus kinematic viscosity of the primordial gas are all well known it is easy to compute the fragmentation masses to be that of protogalaxies composed almost entirely of $10^{24} - 10^{25}$ kg planets in million-solar-mass $10^{36}$ kg (PGC) clumps [4]. The NBDM diffuses to diffusive Schwarz scales $L_{SD} = \left(D^2 / \rho G\right)^{1/4}$ much larger than $L_N$ scale protogalaxies, where $D$ is the NBDM diffusivity and $D \gg \nu$. The rogue-planet prediction of HGD was promptly and independently predicted by the Schild 1996 interpretation of his quasar microlensing observations [14].

The HGD prediction of $\mathrm{Re}_\lambda \sim 100$ weak turbulence in the plasma epoch is supported by statistical studies of cosmic microwave background (CMB) temperature anisotropy fine structure compared to atmospheric, laboratory and numerical simulation turbulence values [19-21]. Plasma turbulence imposes a referred direction to the massive plasma objects formed by gravitational instability in the plasma at length scales that reflect fossil temperature turbulence of the big bang [6]. Baroclinic torques produce vorticity at rate $\partial \vec{\omega} / \partial t = \nabla \rho \times \nabla p / \rho^2$ at the boundary of protosupercluster voids as shown in Fig. 1e [11]. Because $\nabla \rho$ is constant over fossil turbulence Batchelor scales of the big bang turbulence at strong force freeze-out stretched by inflation, HGD explains observations of quasar polarization matching the direction of the Axis of Evil to Gpc scales approaching the present horizon scale $L_H = ct$ [22]. This direction on the cosmic sphere is right ascension RA = $202^\circ$, declination $\delta = 25^\circ$, which matches the 2-4-8-16 directions of CMB spherical harmonics [23] and galaxy spins to supercluster 30 Mpc scales [24].

## 3. Observations

Figure 2 (top) shows the Tadpole (VV29, UGC 10214) galaxy merger system imaged by the HST/ACS camera, compared to a Keck Telescope spectroscopic study (bottom) by Tran et al. 2003 [16]. The galaxy dark matter clearly consists of PGCs since the spectroscopy proves the YGCs were formed in place in the galaxy halo and not ejected as a collisionless tidal tail [17]. Quasar microlensing [14] suggests the dark matter PGCs must be composed of frozen planets in metastable equilibrium, as predicted by HGD [4].

Figure 2 (bottom) shows a linear trail of YGCs pointing precisely to the frictional spiral merger of the small blue galaxy VV29c that is embedded in the accretion disk of VV29a, Fig. 2 (top). With tidal agitation from the merger, the planets undergo an accretional cascade to larger and larger planets, and finally form stars within the PGCs that are stretched away by tidal forces to become field stars in the observed star-wakes and dust-wakes. All galaxies originate as protogalaxies at 0.03 Mpc scales $L_N$ reflecting viscous-gravitational fragmentation of weakly turbulent plasma just before its transition to gas. A core of 13.7 Gyr old stars at scale $L_N$ persists in most if not all galaxies bound by PGC-viscosity of its remaining PGC dark matter. Most of the PGC mass diffuses out of the protogalaxy core to form the galaxy dark matter halo, observed to extend to a diameter of 0.3 Mpc in Tadpole, Fig. 1 (top). A more detailed discussion is found elsewhere [18], and in the present volume.

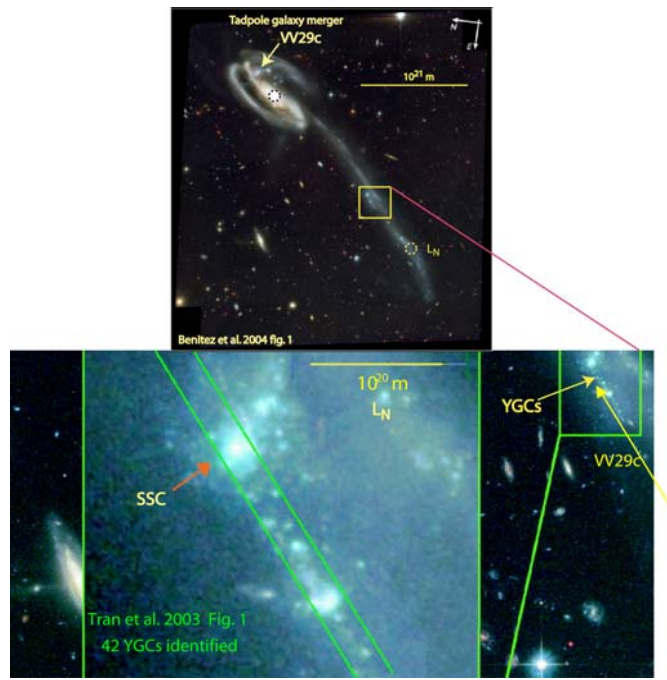

Fig. 2. Tadpole galaxy merger system illustrating the size of the baryonic dark matter system surrounding the central galaxy VV29a and the frictional spiral merger of galaxy VV29c leaving a trail of young-globular-star-clusters (YGCs)

Figure 3 shows a planetary nebula in the Large Magellanic Cloud (LMC) claimed from a recent brightness episode to be on the verge of a SNe Ia event [25], where the central white dwarf and companion (or possibly just a white dwarf and a JPP accretion disk) have ejected several solar masses of matter in the bright clumps observed. The HGD interpretation is very different [10]. From HGD the bright clumps are formed in place from clumped baryonic-dark-matter frozen planets termed JPPs (Jovian PFP Planets of all sizes form by gassy binary accretional mergers of PFPs and their growing daughters), where some of the multi-Jupiter-mass clumps (globulettes [26]) are accreted by the star, and none are ejected. As the JPPs are accreted the star shrinks, its mass and density increase, and its spin rate increases producing a powerful plasma beam that evaporates the frozen gas planets it encounters.

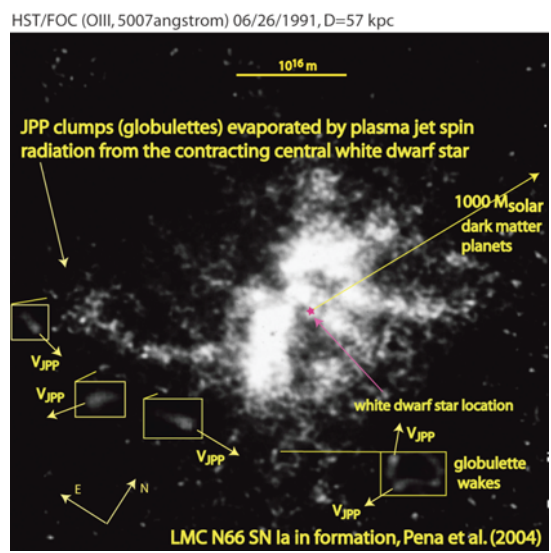

Fig. 3. Planetary nebula LMC N66 suggested by Pena et al. 2004 as a Supernova Ia in formation [25]. From HGD the HST/FOC image reveals globulette clumps [26] of dark matter planets evaporated by the plasma jet of the rapidly spin-ning, contracting, white dwarf star at the center, burdened by the rain of accreting, evaporating, planets.

Bright wakes can be seen for JPP-PFP globulettes (multi-Jupiter mass planet clumps) in the Fig. 3

magnified images (boxes) showing the objects are moving in random directions with speeds $V_{JPP}$. Presumably the globulette speed determines the rate of accretion by the central star and the rate of its radial mixing of thermonuclear products [10]. Slow speeds will reduce the size of the JPPs and their clumps, and increase the probability that the central star will die quietly as a helium white dwarf with mass < 1.4 solar. Moderate JPP accretion rates may fail to mix away the carbon core giving a supernova Ia at the Chandresekhar critical mass 1.44 solar. Stronger mixing and accretion may permit an iron core and a supernova II event. Even stronger accretion rates lead to superstars or black holes. Within the PNe size there should be more than a thousand solar masses of dark matter planets, as shown by the arrow toward upper right.

Figure 4 shows the Helix planetary nebula. Helix is much closer (209 pc) and presumably less strongly agitated by tidal forces from other objects than the LMC N66 PNe of Fig. 3.

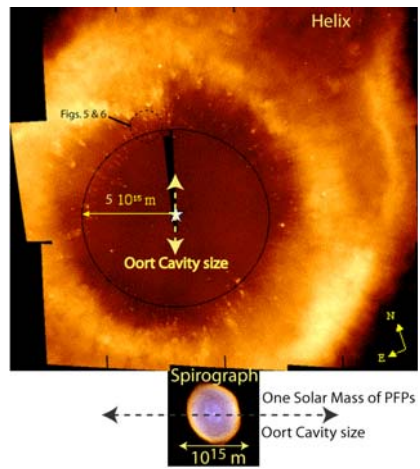

Fig. 4. Helix planetary nebula (top) showing numerous dark matter planets evaporated by the central white dwarf at the outer boundary of the Oort cavity left by the accretion of PFPs to form the star within the PGC. Spirograph PNe (bottom) is young and still growing within its Oort cavity, where apparently no dark matter planets remain unevaporated. Detailed images of Helix dark matter planets are shown in Fig. 5 and Fig. 6 at the indicated location north of the central star.

Thousands of evaporating gas planets can be seen in Fig. 4 (and Fig. 5 and Fig. 6 close-up) HST images. The planets have spacing consistent with the fossil density from the time of first structure at 30,000 years after the big bang; that is, $\rho_0 \sim 10^{-17}$ kg m$^{-3}$.

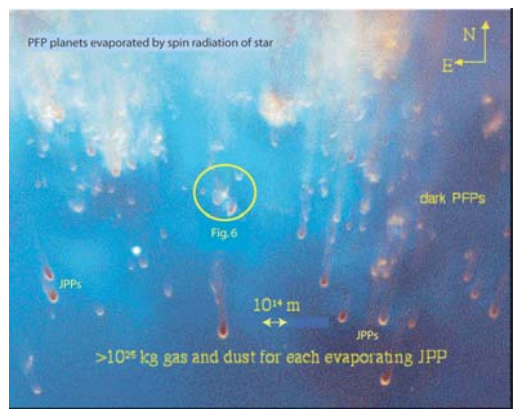

Fig. 5. Close-up image of the region north of the central star shown by the dashed circle in Fig. 4. The central white dwarf spins rapidly because it has shrunk to a density of order $10^{10}$ kg m$^{-3}$ from the mass of accreted planets. A bipolar plasma beam irradiates and evaporates dark matter planets at the edge of the Oort cavity, creating the planetary nebula.

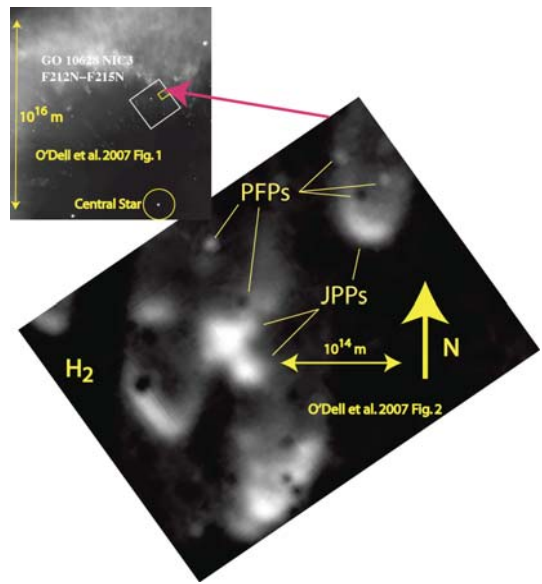

Fig. 6. Detail of location north of central white dwarf star in Helix PNe [28], showing evaporating PFP planets as well as large JPP planets and their atmospheres that can dim a SNe Ia event if it is along the line of sight, Fig. 8.

The mass of white-dwarf-precursor stars are vastly overestimated by ignorance of the existence of dark matter planets. Without PFPs, standard PNe models assume the mass of the nebula originates in the original star and is somehow ejected as clumps and superwinds. Figure 6 shows close-up images of the JPP and partially evaporated PFP planets, illuminated by the spin-radiation of the central star.

Figure 7 shows estimates of the final mass of stars $M_{Final} \leq 1.3 M_{sun} \leq M_{Pulsar} \leq M_{SNeIa}$ compared to the initial mass $M_{PNE}$ where the PNe mass is estimated from the nebulae brightness assuming all of the nebula mass originates in the star and none from dark matter planets.

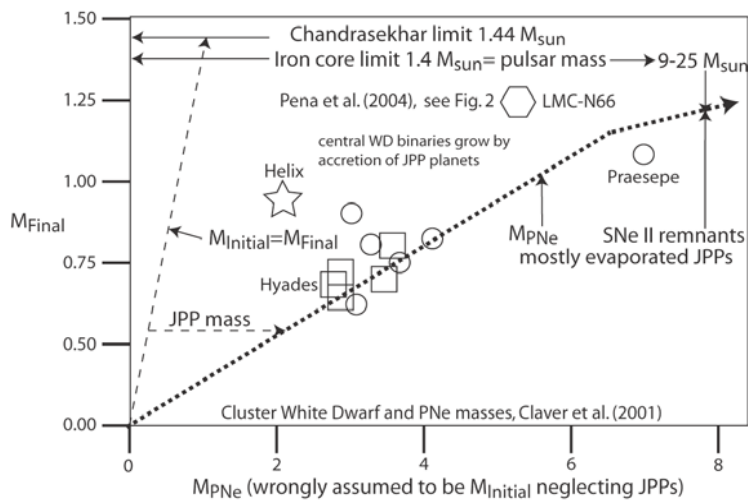

Fig. 7. Star masses compared to assumed PNe and SNe initial star masses neglecting dark matter planet effects such as JPP atmosphere brightness and the mass of evaporated dust and gas [10].

We see from Fig. 7 that final star masses are much smaller than initial masses by standard models of PNe and SNe. Pulsar masses $M_{Pulsar} = 1.4 M_{sun}$ are less than the Chandrasekhar white dwarf limit $M_{Chandrasekhar} = 1.44 M_{sun}$.

Figure 8 shows that an alternative to dark energy is dark matter planets. Fig. 8 (top) is dimness of SNe Ia events versus redshift z for uniform-grey-dust, dark-energy (nonlinear-grey-dust) and no-dark-energy models. The uniform-grey-dust model fails at large z. Dark matter planets provide a nonlinear-grey-dust effect if SNe Ia events take place in PNe surroundings such as the Helix, Fig. 8 (bottom) where spin-radiation of a white dwarf creates JPP atmospheres that may dim or not dim depending on the line of sight.

The brightest SNe Ia events agree with the no-dark-energy curve of Fig. 8 (top) and can be interpreted as lines of sight that do not intersect dense dark matter planet atmospheres. A similar interpretation is given to the Sandage 2006 [27] SNe Ia global Cepheid Hubble-Constant $H_0 = 62.3$ km s$^{-1}$ Mpc$^{-1}$ estimates that disagree with the WMAP $H_0 = 71$ km s$^{-1}$ Mpc$^{-1}$ value and estimate of the age of the universe T to be 15.9 Gyr rather than the CMB value of 13.7 Gyr. Taking the least dim SNe Ia values measured to be correct removes the systematic error of dark matter planet atmosphere dimming (non-linear grey dust), so discrepancies in T and $H_0$ are removed.

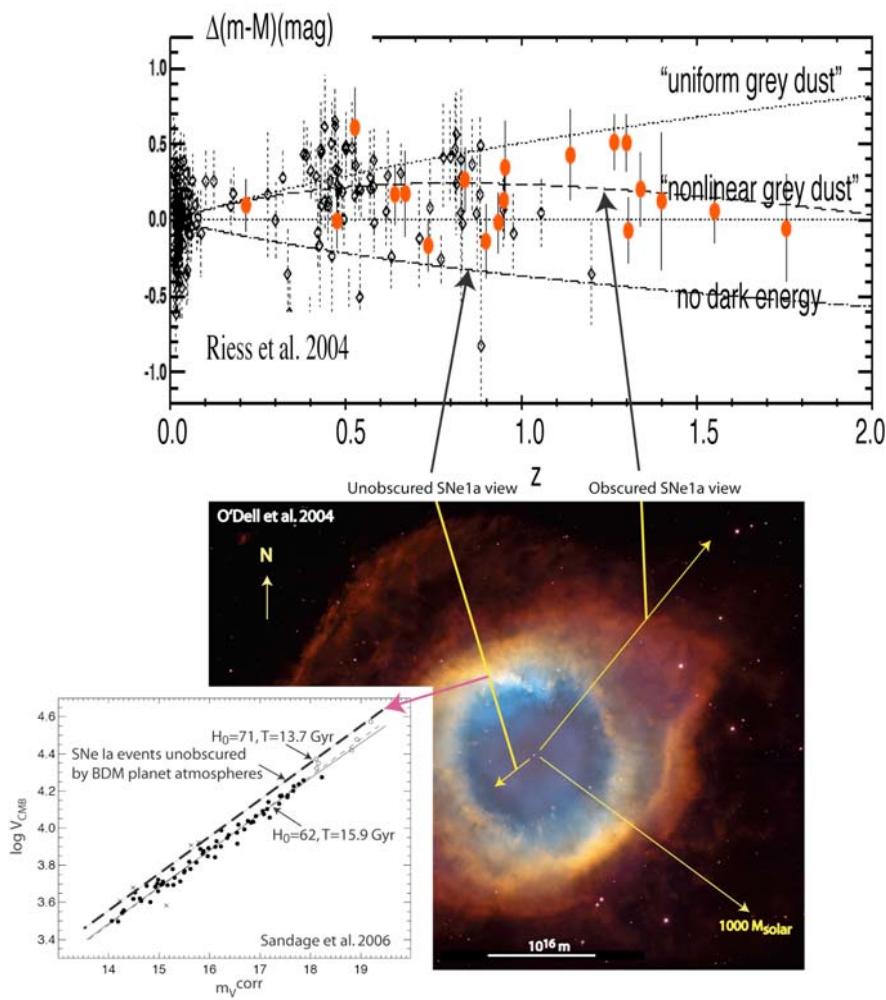

Fig. 8. Helix planetary nebula showing the effect of JPP planetary atmospheres along the line of sight to SNe Ia events is to produce a systematic dimming error that can masquerade both as dark energy (top) or as increased Hubble constants and ages of the universe (left insert).

## 4. Discussion

An accumulation of evidence in a variety of frequency bands from a variety of very high resolution and highly sensitive modern telescopes leaves little doubt that the dark matter of galaxies is primordial planets in proto-globular-star-cluster clumps, as predicted from HGD by Gibson 1996 [4] and inferred from quasar microlensing by Schild 1996 [14]. All stars form from these planets so all star models and planetary nebulae models must be revised to take the effects of planets and their brightness and dimness effects into account. Turbulence produces post-turbulence (fossil turbulence) with structure in patterns that preserve evidence of previous events such as big bang turbulence and plasma epoch turbulence. Post-turbulence perturbations [6] guide the evolution of all subsequent gravitational structures. Numerous fatal flaws in the standard ΛCDMHC cosmology have appeared that can be traced to inappropriate and outdated fluid mechanical assumptions [15] that can be corrected by HGD [4-14].

## 5. Conclusions

Dark matter planet dimming errors account for the SNe Ia overestimate (T=15.9 Gyr) of the age of the universe [27] and dark energy dimming of SNe Ia events, as described by Fig. 8. Dark energy is an unnecessary and incorrect hypothesis from HGD. Thus we need not modify any physical laws nor predict the end of cosmology because evidence of cosmological beginnings are being swept out of sight by an accelerating vacuum-antigravity-powered expansion of the universe [29, 30].

From HGD and the second law of thermodynamics, the frictional non-adiabatic big bang turbulence beginning of the universe [7] implies the universe is closed, not open. Because the big bang was extremely hot, the big bang turbulence produced little entropy so the departures from a flat universe should be small, as observed. Fossil Batchelor scale density turbulence explains the Gpc scales of quasar polarization vectors in alignment with the CMB axis of evil [22, 24, 11], as shown by the dashed circles in Fig. 1.

Star formation models and planetary nebulae formation models must be corrected to account for the effects of dark matter planets [10]. Initial star masses have been vastly overestimated from the brightness of evaporating planets formed around dying central stars by spin radiation, as shown by Fig. 7. Red giant and asymptotic giant branch (AGB) excursions of brightness on the Hertzsprung-Russell color-magnitude diagram for star evolution must be reexamined to account for the fact that all stars are formed and grow by the accretion of planets that have first order effects on both the color and magnitude of stars as they evolve. Planetary nebulae are not dust clouds ejected by AGB superwind events but are partially evaporated and brightly illuminated dark matter planets, as shown by Figs. 3-6, and Fig. 8.